\documentclass{article}
\usepackage{spconf,amsmath,graphicx}
\usepackage[table,xcdraw]{xcolor}
\usepackage[permil]{overpic}

\usepackage{hyperref}
\hypersetup{
    colorlinks=true,
    linkcolor=blue,
    urlcolor=blue,
    }
\newcommand{\vect}[1]{\boldsymbol{\mathrm{#1}}}

\newcommand{\norm}[1]{\left\lVert#1\right\rVert}

\title{Universal speaker recognition encoders for different speech segments duration}
\name{\begin{tabular}{c} Sergey Novoselov$^{1, 2}$,\quad Vladimir Volokhov$^{1, 2}$, \quad Galina Lavrentyeva$^{1}$ \quad \\\end{tabular}}

\address{
$^1$ITMO University, St.Petersburg, Russia \hspace{0.8pt}
$^2$ STC Ltd., St.Petersburg, Russia \hspace{0.8pt}
\\
\texttt{\{sanovoselov,vavolokhov,gmlavrenteva\}@itmo.ru}
}

\begin{document}
%
\maketitle
\begin{abstract}
Creating universal speaker encoders which are robust for different acoustic and speech duration conditions is a big challenge today. According to our observations systems trained on short speech segments are optimal for short phrase speaker verification and systems trained on long segments are superior for long segments verification. A system trained simultaneously on pooled short and long speech segments does not give optimal verification results and usually degrades both for short and long segments. This paper addresses the problem of creating universal speaker encoders for different speech segments duration. We describe our simple recipe for training universal speaker encoder for any type of selected neural network architecture. According to our evaluation results of wav2vec-TDNN based systems obtained for NIST SRE and VoxCeleb1 benchmarks the proposed universal encoder provides speaker verification improvements in case of different enrollment and test speech segment duration. The key feature of the proposed encoder is that it has the same inference time as the selected neural network architecture.  
\end{abstract}
\section{Introduction}
Speaker recognition (SR) is a rapidly growing field that constantly improves its state-of-the-art performance every year. These significant improvements over the last years are mainly determined by the use of deep neural networks, increasing amount of available data and pretrained models.
State-of-the-art \cite{Zeinali2019, garcia2020magneto, gusev2020deep, lee2019nec, evalplan2021nist} SR systems today are based on very deep neural networks like \textit{ResNets}, \textit{ECAPA-TDNNs} \cite{Zeinali2019, garcia2020magneto, gusev2020deep, lee2019nec, evalplan2021nist} and wav2vec \cite{fan2020w2v_speaker, vaessen2021fine, lavrentyeva22_odyssey} or WavLM \cite{chen2021wavlm} like models. The last results obtained for wav2vec based architectures marked a new stage of the development of speaker verification systems based on self-supervised model pre-training. 
However driven by the industry demand this field faces the problems of increasing computational complexity, difficulties in determining the conditions of use and, as a consequence, lack of target data. But what is even more challenging: the demand of universal systems, robust along different domains in terms of channel, noise and speech duration variations.
The mismatch of these factors for enrollment and test trials can dramatically affect the quality of SR systems. 

This paper address the issue of universal speaker recognition system in the context of speech duration mismatch problem.
According to our observations the systems trained on short speech segments are optimal for short phrase speaker verification \cite{gusev2020deep} and systems trained on long segments show better performance for verification on long segments. To deal with these problem the most common approach is to pool short and long speech segments together and train the system simultaneously for all conditions. Unfortunately, such systems does not provide optimal results and usually degrades both for short and long segment cases. The cause for that lies in different lexical variability, presented in short and long speech segments (and the way neural network are trained to deal with that).

Combining multiple systems based on different classifiers trained for specific conditions reduces speaker verification error rate with short utterances. However that increase the computational cost of such systems.

What if we could model a common embedding space, where embeddings from different SR systems trained for specific conditions can be compared using some simple scoring, like cosine.

This paper present a novel approach to do it based on cl-embeddings \cite{lavrentyeva22_odyssey}. The similar idea was considered in \cite{asymmetric} for universal space modelling using specific losses. Here, we do not consider any specific tools during the training process and the proposed solution is based only on the already trained encoders.
We observe the lack of the generalization of SR encoders for long and short speech segments even if one uses huge transformers-based encoders like large wav2vec-TDNN models \cite{lavrentyeva22_odyssey}.
The contributions of our paper are as follows:
\begin{itemize}
\setlength\itemsep{-0.3em}
    \item We explore the recent concept of cl-embeddings space which allows to achieve better performance while leaving all the positive properties of classic embeddings space (small embedding size, cosine scoring) and demonstrate that this approach also allows to compare embeddings obtained by two different encoders trained on the same set of speakers. 
    \item We propose a novel universal speaker extractor approach based on cl-embedding concept considering the long and short speech segments problem. We demonstrate that the quality of the proposed approach is comparable to, and in some cases better than, the quality of systems trained for specific conditions. 
    \item We confirmed the effectiveness of the proposed approach for the state-of-the-art wav2vec 2.0 based model on the wide range of available datasets.
    
\end{itemize}

\section{Deep speaker encoders}
\label{sec:encoders}

The conventional deep neural network based solution for extracting utterance-level speaker embeddings consists of three blocks: an encoder network for extracting frame-level representations from the acoustic features, pooling layer that converts variable-length frame-level features into one fixed-dimensional vector and a feed forward classification network that processes the pooling vector to produce speaker class posterior. The role of the encoder network can be taken by a neural network of any type.

In this work we experimented with strong wav2vec 2.0 based SR systems introduced recently in \cite{lavrentyeva22_odyssey}.   



\textit{Wav2vec-TDNN encoder network.}
\textbf{Wav2vec 2.0} model is a powerful transformer-based model developed for ASR tasks. It takes raw speech signals (16 kHz) as input and incorporates a multi-head attention mechanism on the deep layers. The key aspect of training such a model is Contrastive Predicting Coding \cite{oord2018representation} self-supervised pretraining scheme. It was shown in \cite{baevski2020wav2vec} that wav2vec 2.0 model pretrained on large amounts of diverse and unlabelled data can be successfully fine-tuned to specific low resource ASR tasks. 

As an effective wav2vec 2.0 backend we applied two TDNN layers (the 1st with ReLU activation), statistic pooling layer to pool time series to a single vector, maxout linear layer \cite{gusev2020deep, novoselov2018deep} to obtain speaker embedding. We used an AAM-Softmax-based linear classification layer to fine-tune the extractor. 

Large multi-lingual wav2vec 2.0  \textit{XLS-R\_1B} \footnotemark[1] model provided by Facebook \cite{ott2019fairseq, babu2021xlsr} were used as starting points for the extractors fine-tuning. We named the corresponding speaker embedding extractors as w2v-tdnn.


In addition, following our experience described in \cite{lavrentyeva22_odyssey} we performed speaker verification for considered systems using the cl-embeddings.

\footnotetext[1]{\href{https://github.com/pytorch/fairseq/tree/main/examples/wav2vec/xlsr}{https://github.com/pytorch/fairseq/tree/main/examples/wav2vec/xlsr}}

\section{Common embedding space}
\label{sec:space}
\subsection{Cl-embeddings}
The standard speaker embeddings are extracted from the segment level of the neural network trained using the angular margin loss. These speaker embeddings are further discriminated using cosine similarity:
\begin{equation}
\label{eq:cos}
\mathcal{S(\vect{e_1},\vect{e_2})} = \dfrac{\vect{e_1}^T\vect{e_2}}{{\norm{\vect{e_1}}}{\norm{\vect{e_2}}}},
\end{equation}
where $(\vect{e_1}, \vect{e_2})$ are speaker embedding vectors.

Apart from the typical approach, recently proposed cl-embeddings  \cite{lavrentyeva22_odyssey} are extracted from the last classification linear layer. 
It was revealed that using cl-embeddings instead of the conventional embeddings leads to valuable performance improvement for an already trained system without any additional operations.
Indeed, since the encoder was trained to separate speaker voices, its linear layer contains a huge amount of informative parameters that can be used for speaker discrimination not only during the training process but on the evaluation step as well. 
Systems based on cl-embeddings achieved state-of-the-art performance in NIST SRE 2021 \cite{lavrentyeva22_odyssey}.


Recent paper \cite{themos} showed that the problem of high dimension can be efficiently solved by Cholesky (or other equivalent) decomposition of symmetric positive (semi-)definite matrices.

Let us consider $l$-dimentional vectors $\vect{e_1}$ and $\vect{e_2}$ - the examples of conventional pre-last layer embeddings and $N_{class}$-dimension vectors $\vect{c}_1$ and $\vect{c}_2$ are corresponding cl-embeddins:
\begin{equation}
 \vect{c}_{i} = \mathbf{W}_{class}^T\vect{e}_{i}, i \in [1, 2]\\
\label{cl-embs}
\end{equation}
 Here $\mathbf{W}_{class}$ is the classification linear layer matrix with size $(l$ x $N_{class})$.
$L_2$ - norm in the cl-embedding space can be computed as:
\begin{equation}
 ||\vect{c}_{i}||^2 = \vect{c}_{i}^T\vect{c}_{i} = \vect{e}_{i}^T \mathbf{A}\vect{e}_{i} = \vect{y}_{i}^T\vect{y}_{i}\\
\label{eq:l2_norm}
\end{equation}
where $\mathbf{A} = \mathbf{W}_{class}\mathbf{W}_{class}^T$ and $\vect{y}_{i} = \mathbf{L}^T\vect{e}_{i}$ and $\mathbf{A} = \mathbf{L}\mathbf{L}^T$ is a Cholesky decomposition of the matrix $\mathbf{A}$. At the same time:

\begin{equation}
 \vect{c}_1^T\vect{c}_2 = \vect{e}_{1}^T \mathbf{A}\vect{e}_{2} = \vect{y}_{1}^T\vect{y}_{2}\\
\label{eq:dot_prod}
\end{equation}
 Eq. \ref{eq:l2_norm} and \ref{eq:dot_prod} prove that $\mathcal{S(\vect{c_1},\vect{c_2})} = \mathcal{S(\vect{y_1},\vect{y_2})}$,
where $\vect{y}_{1}$ and $\vect{y}_{2}$ live in low dimentional space like $\vect{e}_{(1,2)}$.

It should be noted that one can use eigendecomposition of $\mathbf{A}$ with dimentional reduction option instead of Cholesky: $\mathbf{A} = \mathbf{U}^T \cdot \Lambda \cdot \mathbf{U} = (\Lambda ^{1/2} \cdot \mathbf{U})^T \cdot (\Lambda ^{1/2} \cdot \mathbf{U})$ where $\Lambda$ is the diagonal matrix whose diagonal elements are the corresponding eigenvalues and $\mathbf{U}$ is a matrix of eigenvectors.

\subsection{Fusion space}
As soon as we have mapping to the same posterior space the proposed cl-embeddings can be used for embedding level fusion in case we have common trained dataset or representative subset.
Moreover, one can use different fusion weights while embedding level fusion in enrollment and test embeddings inference processes, since they stay in one space of speaker posteriors logits of the same classes.
Additionally, appendix B in \cite{themos} confirmed that this fusion can be performed in the reduced cl-embedding space. 

In case of embedding fusion of two different speaker encoders having l-dimentional embeddings $\vect{e}^1$ and $\vect{e}^2$ and with $\mathbf{W}_{class}^1$ and $\mathbf{W}_{class}^2$  classification heads with the same set of train classes one deals with the concatenated projection fusion matrix $\mathbf{W}_{class}^f = [\mathbf{W}_{class}^1, \mathbf{W}_{class}^2]$ with size $(2l$ x $N_{class})$
and matrix $\mathbf{A}^f = \mathbf{W}_{class}^f(\mathbf{W}_{class}^f)^T$ that can be decomposed as $\mathbf{A}^f = \mathbf{L}^f(\mathbf{L}^f)^T$. Decomposition matrix $\mathbf{L}^f$ can have size of $(2l$ x $l^r)$ in case of using eigenvectors decomposition with dimention reduction from $2l$ to $l^r$. One can also deconcatenate $\mathbf{L}^f = [\mathbf{L}_1^f, \mathbf{L}_2^f]$. This means that obtained reduced fusion cl-embedding space $\vect{y}^f$ will have size of $l^r$ and one can perform fusion in this space simply by using $\vect{y}^f = w_1\vect{y}_1^f + w_2\vect{y}_2^f$ where $\vect{y}_{1}^f = (\mathbf{L}_1^f)^T\vect{e}^{1}$ and $\vect{y}_{2}^f = (\mathbf{L}_2^f)^T\vect{e}^{2}$ and $w_1$, $w_2$ - fusion weights. In our experiments we observed that fusion embedding space can be reduced to size of 200 - 256 without the loss of the generalization.

 It is easy to show that $\mathcal{S(\vect{c_1},\vect{c_2})} \approx \mathcal{S(\vect{y_1^f},\vect{y_2^f})}$ for different encoders as soon as they both map features into one speaker embedding space. Thus such linear projectors ${L}_1^f$ and ${L}_2^f$ allows us to obtain common low dimentional embedding space $\vect{y}^f$ from conventional embeddings $\vect{e}_1$ and $\vect{e}_2$ for two different speaker encoders.


The capability to obtain common embedding space for different encoders led us to the idea of a universal speaker encoder.


\section{Universal speaker encoders}
\label{sec:uni-encoders}
\begin{figure}
    \centering
    \begin{overpic}[width=0.45\textwidth]{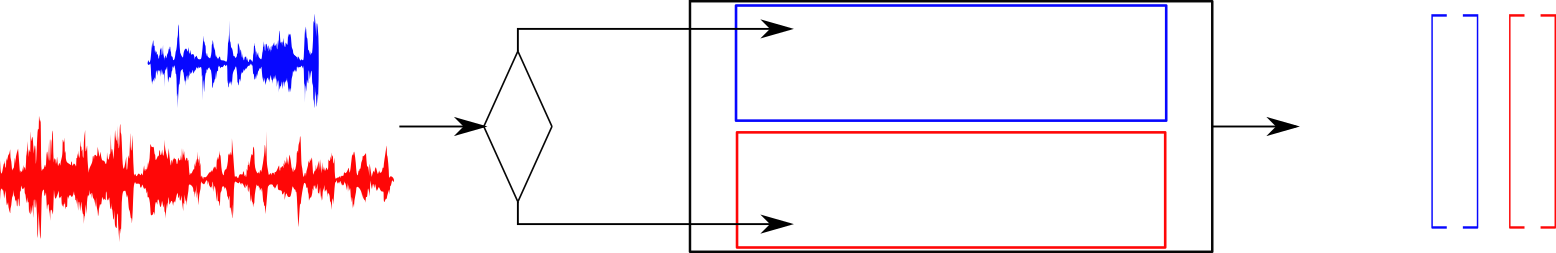}%
\put(350,155){\small short}%
\put(350,-5){\small long}%
\put(550,110){\small Encoder$_1$}%
\put(550,30){\small Encoder$_2$}%
\put(860,80){$S\left(\quad,\:\:\;\right)$}%

\end{overpic}
    \caption{Universal speaker encoder scheme for different speech segments duration}
    \label{fig:uni_scheme}
\end{figure}
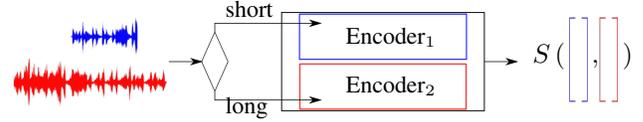

Let us consider the speaker verification scenario when enrollment and test speech segments correspond to different recording conditions. For example, enrollment segment is recorded in the controlled acoustic conditions while test segments is recorded in-the-wild. Or, another example: enrollment speech segment is much longer than the test segment. Suppose one managed to fine-tune two encoders for different conditions: first encoder for long duration segments and second encoder for short duration segments. Taking into account the ideas from Section \ref{sec:space} about common low dimensional embedding space we can propose universal speaker encoder. It is based on the selection of the appropriate encoder for each specific input segment according to its duration. In other words it maps different speech segments using different encoders to the same embedding space where they can be compared. 
Figure \ref{fig:uni_scheme} demonstrates the scheme of the proposed universal speaker encoder for different speech segment duration. It is represented by two base encoders: Encoder$_1$ is fine-tuned for short segments and Encoder$_2$ is adapted for long segments. Thus the universal encoder doubles the size of the base encoder but keeps its inference time. The universal encoder works as follows: if it has a short input speech segment then it selects Encoder$_1$ to extract speaker embedding $\vect{y}_{1}^f$ and if it deals with a long speech segment as input than Encoder$_2$ is used for $\vect{y}_{2}^f$ speaker embedding extraction. It is important that obtained embeddings represent common embedding space that ensures the capability of such embeddings comparisons using simple cosine scoring (see Section \ref{sec:space}).

\begin{table*}[h]
\centering
\caption{Speaker recognition evaluations on different test protocols for \textit{full} enrollment speech segment duration and different test speech segment duration in terms of \textit{EER {[}\%{]}  / minDCF(0.05).}}

\label{tab:main_table_full_enr}
\scalebox{0.74}{
\centerline{
\begin{tabular}{lllllllllllllllllllll}
\cline{1-1} \cline{3-6} \cline{8-11} \cline{13-16} \cline{18-21}
\multicolumn{1}{|l|}{\cellcolor[HTML]{FFFFC7}\textbf{Encoder name}} &
  \multicolumn{1}{l|}{\textbf{}} &
  \multicolumn{4}{c|}{\cellcolor[HTML]{FFFFC7}\textbf{w2v-tdnn-long}} &
  \multicolumn{1}{c|}{\textbf{}} &
  \multicolumn{4}{c|}{\cellcolor[HTML]{FFFFC7}\textbf{w2v-tdnn-short}} &
  \multicolumn{1}{c|}{\textbf{}} &
  \multicolumn{4}{c|}{\cellcolor[HTML]{FFFFC7}\textbf{w2v-tdnn-pooled}} &
  \multicolumn{1}{c|}{\textbf{}} &
  \multicolumn{4}{c|}{\cellcolor[HTML]{FFFFC7}\textbf{w2v-tdnn-uni}} \\ \cline{1-1} \cline{3-6} \cline{8-11} \cline{13-16} \cline{18-21} 
\multicolumn{1}{|l|}{\cellcolor[HTML]{FFFFC7}\textbf{Test duration, {[}sec{]}}} &
  \multicolumn{1}{l|}{\textbf{}} &
  \multicolumn{1}{c|}{\cellcolor[HTML]{FFFFC7}\textbf{2}} &
  \multicolumn{1}{c|}{\cellcolor[HTML]{FFFFC7}\textbf{3}} &
  \multicolumn{1}{c|}{\cellcolor[HTML]{FFFFC7}\textbf{4}} &
  \multicolumn{1}{c|}{\cellcolor[HTML]{FFFFC7}\textbf{full}} &
  \multicolumn{1}{c|}{\textbf{}} &
  \multicolumn{1}{c|}{\cellcolor[HTML]{FFFFC7}\textbf{2}} &
  \multicolumn{1}{c|}{\cellcolor[HTML]{FFFFC7}\textbf{3}} &
  \multicolumn{1}{c|}{\cellcolor[HTML]{FFFFC7}\textbf{4}} &
  \multicolumn{1}{c|}{\cellcolor[HTML]{FFFFC7}\textbf{full}} &
  \multicolumn{1}{c|}{\textbf{}} &
  \multicolumn{1}{c|}{\cellcolor[HTML]{FFFFC7}\textbf{2}} &
  \multicolumn{1}{c|}{\cellcolor[HTML]{FFFFC7}\textbf{3}} &
  \multicolumn{1}{c|}{\cellcolor[HTML]{FFFFC7}\textbf{4}} &
  \multicolumn{1}{c|}{\cellcolor[HTML]{FFFFC7}\textbf{full}} &
  \multicolumn{1}{c|}{\textbf{}} &
  \multicolumn{1}{c|}{\cellcolor[HTML]{FFFFC7}\textbf{2}} &
  \multicolumn{1}{c|}{\cellcolor[HTML]{FFFFC7}\textbf{3}} &
  \multicolumn{1}{c|}{\cellcolor[HTML]{FFFFC7}\textbf{4}} &
  \multicolumn{1}{c|}{\cellcolor[HTML]{FFFFC7}\textbf{full}} \\ \cline{1-1} \cline{3-6} \cline{8-11} \cline{13-16} \cline{18-21} 
\textbf{Dataset} &
  \textbf{} &
  \multicolumn{19}{c}{\textbf{}} \\ \cline{1-1} \cline{3-6} \cline{8-11} \cline{13-16} \cline{18-21} 
\multicolumn{1}{|l|}{\cellcolor[HTML]{FFFFC7}VC1-O (cleaned)} &
  \multicolumn{1}{l|}{} &
  \multicolumn{1}{l|}{\begin{tabular}[c]{@{}l@{}}6.14 \\ /0.37\end{tabular}} &
  \multicolumn{1}{l|}{\begin{tabular}[c]{@{}l@{}}2.93  \\ /0.2\end{tabular}} &
  \multicolumn{1}{l|}{\begin{tabular}[c]{@{}l@{}}1.65  \\  /0.12\end{tabular}} &
  \multicolumn{1}{l|}{\begin{tabular}[c]{@{}l@{}}0.95  \\  /0.06\end{tabular}} &
  \multicolumn{1}{l|}{} &
  \multicolumn{1}{l|}{\begin{tabular}[c]{@{}l@{}}3.35 \\ /0.21\end{tabular}} &
  \multicolumn{1}{l|}{\begin{tabular}[c]{@{}l@{}}1.88  \\  /0.13\end{tabular}} &
  \multicolumn{1}{l|}{\begin{tabular}[c]{@{}l@{}}1.39  \\  /0.09\end{tabular}} &
  \multicolumn{1}{l|}{\begin{tabular}[c]{@{}l@{}}1.02 \\  /0.07\end{tabular}} &
  \multicolumn{1}{l|}{} &
  \multicolumn{1}{l|}{\begin{tabular}[c]{@{}l@{}}4.59 \\ /0.3\end{tabular}} &
  \multicolumn{1}{l|}{\begin{tabular}[c]{@{}l@{}}2.37   \\ /0.16\end{tabular}} &
  \multicolumn{1}{l|}{\textbf{\begin{tabular}[c]{@{}l@{}}1.43   \\ /0.11\end{tabular}}} &
  \multicolumn{1}{l|}{\textbf{\begin{tabular}[c]{@{}l@{}}0.94   \\ /0.06\end{tabular}}} &
  \multicolumn{1}{l|}{} &
  \multicolumn{1}{l|}{\textbf{\begin{tabular}[c]{@{}l@{}}3.35 \\ /0.21\end{tabular}}} &
  \multicolumn{1}{l|}{\textbf{\begin{tabular}[c]{@{}l@{}}1.88   \\ /0.13\end{tabular}}} &
  \multicolumn{1}{l|}{\begin{tabular}[c]{@{}l@{}}1.65   \\ /0.12\end{tabular}} &
  \multicolumn{1}{l|}{\begin{tabular}[c]{@{}l@{}}0.96   \\ /0.06\end{tabular}} \\ \cline{1-1} \cline{3-6} \cline{8-11} \cline{13-16} \cline{18-21} 
\multicolumn{1}{|l|}{\cellcolor[HTML]{FFFFC7}SRE'18 dev} &
  \multicolumn{1}{l|}{} &
  \multicolumn{1}{l|}{\begin{tabular}[c]{@{}l@{}}14.03 \\ /0.47\end{tabular}} &
  \multicolumn{1}{l|}{\begin{tabular}[c]{@{}l@{}}10.29   \\ /0.37\end{tabular}} &
  \multicolumn{1}{l|}{\begin{tabular}[c]{@{}l@{}}8.35   \\ /0.3\end{tabular}} &
  \multicolumn{1}{l|}{\begin{tabular}[c]{@{}l@{}}3  \\  /0.08\end{tabular}} &
  \multicolumn{1}{l|}{} &
  \multicolumn{1}{l|}{\begin{tabular}[c]{@{}l@{}}11.92\\  /0.48\end{tabular}} &
  \multicolumn{1}{l|}{\begin{tabular}[c]{@{}l@{}}10.28  \\  /0.4\end{tabular}} &
  \multicolumn{1}{l|}{\begin{tabular}[c]{@{}l@{}}9.35  \\  /0.36\end{tabular}} &
  \multicolumn{1}{l|}{\begin{tabular}[c]{@{}l@{}}5.55  \\  /0.23\end{tabular}} &
  \multicolumn{1}{l|}{} &
  \multicolumn{1}{l|}{\begin{tabular}[c]{@{}l@{}}13.03 \\ /0.47\end{tabular}} &
  \multicolumn{1}{l|}{\begin{tabular}[c]{@{}l@{}}9.95   \\ /0.37\end{tabular}} &
  \multicolumn{1}{l|}{\begin{tabular}[c]{@{}l@{}}8.36   \\ /0.32\end{tabular}} &
  \multicolumn{1}{l|}{\begin{tabular}[c]{@{}l@{}}3.23   \\ /0.12\end{tabular}} &
  \multicolumn{1}{l|}{} &
  \multicolumn{1}{l|}{\textbf{\begin{tabular}[c]{@{}l@{}}11.74 \\ /0.44\end{tabular}}} &
  \multicolumn{1}{l|}{\textbf{\begin{tabular}[c]{@{}l@{}}9.62   \\ /0.36\end{tabular}}} &
  \multicolumn{1}{l|}{\textbf{\begin{tabular}[c]{@{}l@{}}8.35  \\  /0.3\end{tabular}}} &
  \multicolumn{1}{l|}{\textbf{\begin{tabular}[c]{@{}l@{}}3   \\ /0.08\end{tabular}}} \\ \cline{1-1} \cline{3-6} \cline{8-11} \cline{13-16} \cline{18-21} 
\multicolumn{1}{|l|}{\cellcolor[HTML]{FFFFC7}SRE'16 eval} &
  \multicolumn{1}{l|}{} &
  \multicolumn{1}{l|}{\begin{tabular}[c]{@{}l@{}}14.72 \\ /0.6\end{tabular}} &
  \multicolumn{1}{l|}{\begin{tabular}[c]{@{}l@{}}11.17   \\ /0.48\end{tabular}} &
  \multicolumn{1}{l|}{\begin{tabular}[c]{@{}l@{}}9.03   \\ /0.41\end{tabular}} &
  \multicolumn{1}{l|}{\begin{tabular}[c]{@{}l@{}}3.47  \\  /0.18\end{tabular}} &
  \multicolumn{1}{l|}{} &
  \multicolumn{1}{l|}{\begin{tabular}[c]{@{}l@{}}14.68 \\ /0.72\end{tabular}} &
  \multicolumn{1}{l|}{\begin{tabular}[c]{@{}l@{}}12.66  \\  /0.66\end{tabular}} &
  \multicolumn{1}{l|}{\begin{tabular}[c]{@{}l@{}}11.65  \\  /0.62\end{tabular}} &
  \multicolumn{1}{l|}{\begin{tabular}[c]{@{}l@{}}8.75   \\ /0.5\end{tabular}} &
  \multicolumn{1}{l|}{} &
  \multicolumn{1}{l|}{\begin{tabular}[c]{@{}l@{}}14.06 \\ /0.6\end{tabular}} &
  \multicolumn{1}{l|}{\textbf{\begin{tabular}[c]{@{}l@{}}10.9   \\ /0.5\end{tabular}}} &
  \multicolumn{1}{l|}{\begin{tabular}[c]{@{}l@{}}9.17   \\ /0.44\end{tabular}} &
  \multicolumn{1}{l|}{\begin{tabular}[c]{@{}l@{}}4.17   \\ /0.25\end{tabular}} &
  \multicolumn{1}{l|}{} &
  \multicolumn{1}{l|}{\textbf{\begin{tabular}[c]{@{}l@{}}13.62 \\ /0.61\end{tabular}}} &
  \multicolumn{1}{l|}{\begin{tabular}[c]{@{}l@{}}11.08   \\ /0.54\end{tabular}} &
  \multicolumn{1}{l|}{\textbf{\begin{tabular}[c]{@{}l@{}}9.03   \\ /0.41\end{tabular}}} &
  \multicolumn{1}{l|}{\textbf{\begin{tabular}[c]{@{}l@{}}3.47   \\ /0.18\end{tabular}}} \\ \cline{1-1} \cline{3-6} \cline{8-11} \cline{13-16} \cline{18-21} 
\multicolumn{1}{|l|}{\cellcolor[HTML]{FFFFC7}SRE'19 eval} &
  \multicolumn{1}{l|}{} &
  \multicolumn{1}{l|}{\begin{tabular}[c]{@{}l@{}}11.36 \\ /0.4\end{tabular}} &
  \multicolumn{1}{l|}{\begin{tabular}[c]{@{}l@{}}8.28   \\ /0.36\end{tabular}} &
  \multicolumn{1}{l|}{\begin{tabular}[c]{@{}l@{}}6.7  \\  /0.29\end{tabular}} &
  \multicolumn{1}{l|}{\begin{tabular}[c]{@{}l@{}}1.82  \\  /0.11\end{tabular}} &
  \multicolumn{1}{l|}{} &
  \multicolumn{1}{l|}{\begin{tabular}[c]{@{}l@{}}10.57 \\ /0.47\end{tabular}} &
  \multicolumn{1}{l|}{\begin{tabular}[c]{@{}l@{}}8.54  \\  /0.39\end{tabular}} &
  \multicolumn{1}{l|}{\begin{tabular}[c]{@{}l@{}}7.51  \\  /0.35\end{tabular}} &
  \multicolumn{1}{l|}{\begin{tabular}[c]{@{}l@{}}4   \\ /0.24\end{tabular}} &
  \multicolumn{1}{l|}{} &
  \multicolumn{1}{l|}{\begin{tabular}[c]{@{}l@{}}10.73 \\ /0.46\end{tabular}} &
  \multicolumn{1}{l|}{\begin{tabular}[c]{@{}l@{}}8.09   \\ /0.36\end{tabular}} &
  \multicolumn{1}{l|}{\begin{tabular}[c]{@{}l@{}}6.75   \\ /0.31\end{tabular}} &
  \multicolumn{1}{l|}{\begin{tabular}[c]{@{}l@{}}2.31   \\ /0.14\end{tabular}} &
  \multicolumn{1}{l|}{} &
  \multicolumn{1}{l|}{\textbf{\begin{tabular}[c]{@{}l@{}}9.78 \\ /0.43\end{tabular}}} &
  \multicolumn{1}{l|}{\textbf{\begin{tabular}[c]{@{}l@{}}7.61   \\ /0.35\end{tabular}}} &
  \multicolumn{1}{l|}{\textbf{\begin{tabular}[c]{@{}l@{}}6.7   \\ /0.29\end{tabular}}} &
  \multicolumn{1}{l|}{\textbf{\begin{tabular}[c]{@{}l@{}}1.82   \\ /0.11\end{tabular}}} \\ \cline{1-1} \cline{3-6} \cline{8-11} \cline{13-16} \cline{18-21} 
\multicolumn{1}{|l|}{\cellcolor[HTML]{FFFFC7}rus IVR mic} &
  \multicolumn{1}{l|}{} &
  \multicolumn{1}{l|}{\begin{tabular}[c]{@{}l@{}}4.89 \\ /0.26\end{tabular}} &
  \multicolumn{1}{l|}{\begin{tabular}[c]{@{}l@{}}2.97   \\ /0.18\end{tabular}} &
  \multicolumn{1}{l|}{\begin{tabular}[c]{@{}l@{}}2.44   \\ /0.15\end{tabular}} &
  \multicolumn{1}{l|}{\begin{tabular}[c]{@{}l@{}}1.12  \\  /0.08\end{tabular}} &
  \multicolumn{1}{l|}{} &
  \multicolumn{1}{l|}{\begin{tabular}[c]{@{}l@{}}4.39 \\ /0.25\end{tabular}} &
  \multicolumn{1}{l|}{\begin{tabular}[c]{@{}l@{}}3.25  \\  /0.19\end{tabular}} &
  \multicolumn{1}{l|}{\begin{tabular}[c]{@{}l@{}}2.93   \\ /0.18\end{tabular}} &
  \multicolumn{1}{l|}{\begin{tabular}[c]{@{}l@{}}2.06   \\ /0.13\end{tabular}} &
  \multicolumn{1}{l|}{} &
  \multicolumn{1}{l|}{\begin{tabular}[c]{@{}l@{}}4.28 \\ /0.24\end{tabular}} &
  \multicolumn{1}{l|}{\textbf{\begin{tabular}[c]{@{}l@{}}2.79   \\ /0.17\end{tabular}}} &
  \multicolumn{1}{l|}{\textbf{\begin{tabular}[c]{@{}l@{}}2.35   \\ /0.15\end{tabular}}} &
  \multicolumn{1}{l|}{\begin{tabular}[c]{@{}l@{}}1.32   \\ /0.09\end{tabular}} &
  \multicolumn{1}{l|}{} &
  \multicolumn{1}{l|}{\textbf{\begin{tabular}[c]{@{}l@{}}4.05 \\ /0.23\end{tabular}}} &
  \multicolumn{1}{l|}{\begin{tabular}[c]{@{}l@{}}2.87   \\ /0.17\end{tabular}} &
  \multicolumn{1}{l|}{\begin{tabular}[c]{@{}l@{}}2.44   \\ /0.15\end{tabular}} &
  \multicolumn{1}{l|}{\textbf{\begin{tabular}[c]{@{}l@{}}1.12   \\ /0.08\end{tabular}}} \\ \cline{1-1} \cline{3-6} \cline{8-11} \cline{13-16} \cline{18-21} 
\multicolumn{1}{|l|}{\cellcolor[HTML]{FFFFC7}rus IVR phn} &
  \multicolumn{1}{l|}{} &
  \multicolumn{1}{l|}{\begin{tabular}[c]{@{}l@{}}3.94 \\ /0.24\end{tabular}} &
  \multicolumn{1}{l|}{\begin{tabular}[c]{@{}l@{}}2.7   \\ /0.17\end{tabular}} &
  \multicolumn{1}{l|}{\begin{tabular}[c]{@{}l@{}}2.17   \\ /0.14\end{tabular}} &
  \multicolumn{1}{l|}{\begin{tabular}[c]{@{}l@{}}1.01  \\  /0.07\end{tabular}} &
  \multicolumn{1}{l|}{} &
  \multicolumn{1}{l|}{\begin{tabular}[c]{@{}l@{}}4.22 \\ /0.26\end{tabular}} &
  \multicolumn{1}{l|}{\begin{tabular}[c]{@{}l@{}}3.51  \\  /0.21\end{tabular}} &
  \multicolumn{1}{l|}{\begin{tabular}[c]{@{}l@{}}3.18  \\  /0.2\end{tabular}} &
  \multicolumn{1}{l|}{\begin{tabular}[c]{@{}l@{}}2.44   \\ /0.15\end{tabular}} &
  \multicolumn{1}{l|}{} &
  \multicolumn{1}{l|}{\begin{tabular}[c]{@{}l@{}}3.72\\ /0.24\end{tabular}} &
  \multicolumn{1}{l|}{\begin{tabular}[c]{@{}l@{}}2.8   \\ /0.18\end{tabular}} &
  \multicolumn{1}{l|}{\begin{tabular}[c]{@{}l@{}}2.37   \\ /0.15\end{tabular}} &
  \multicolumn{1}{l|}{\begin{tabular}[c]{@{}l@{}}1.4   \\ /0.09\end{tabular}} &
  \multicolumn{1}{l|}{} &
  \multicolumn{1}{l|}{\textbf{\begin{tabular}[c]{@{}l@{}}3.22 \\ /0.22\end{tabular}}} &
  \multicolumn{1}{l|}{\textbf{\begin{tabular}[c]{@{}l@{}}2.6  \\ /0.17\end{tabular}}} &
  \multicolumn{1}{l|}{\textbf{\begin{tabular}[c]{@{}l@{}}2.18   \\ /0.14\end{tabular}}} &
  \multicolumn{1}{l|}{\textbf{\begin{tabular}[c]{@{}l@{}}1.01   \\ /0.07\end{tabular}}} \\ \cline{1-1} \cline{3-6} \cline{8-11} \cline{13-16} \cline{18-21} 
\end{tabular}
}
}
\end{table*}

\begin{table*}[h]
\centering
\caption{Speaker recognition evaluations on different test protocols for \textit{3 second} enrollment speech segment  and different test speech segment duration in terms of \textit{EER {[}\%{]}  / minDCF(0.05).}}

\label{tab:main_table_3sec_enr}
\scalebox{0.73}{
\centerline{
\begin{tabular}{lllllllllllllllllllll}
\cline{1-1} \cline{3-6} \cline{8-11} \cline{13-16} \cline{18-21}
\multicolumn{1}{|l|}{\cellcolor[HTML]{FFFFC7}\textbf{Encoder name}} &
  \multicolumn{1}{l|}{\textbf{}} &
  \multicolumn{4}{c|}{\cellcolor[HTML]{FFFFC7}\textbf{w2v-tdnn-long}} &
  \multicolumn{1}{c|}{\textbf{}} &
  \multicolumn{4}{c|}{\cellcolor[HTML]{FFFFC7}\textbf{w2v-tdnn-short}} &
  \multicolumn{1}{c|}{\textbf{}} &
  \multicolumn{4}{c|}{\cellcolor[HTML]{FFFFC7}\textbf{w2v-tdnn-pooled}} &
  \multicolumn{1}{c|}{\textbf{}} &
  \multicolumn{4}{c|}{\cellcolor[HTML]{FFFFC7}\textbf{w2v-tdnn-uni}} \\ \cline{1-1} \cline{3-6} \cline{8-11} \cline{13-16} \cline{18-21} 
\multicolumn{1}{|l|}{\cellcolor[HTML]{FFFFC7}\textbf{Test duration, {[}sec{]}}} &
  \multicolumn{1}{l|}{\textbf{}} &
  \multicolumn{1}{c|}{\cellcolor[HTML]{FFFFC7}\textbf{2}} &
  \multicolumn{1}{c|}{\cellcolor[HTML]{FFFFC7}\textbf{3}} &
  \multicolumn{1}{c|}{\cellcolor[HTML]{FFFFC7}\textbf{4}} &
  \multicolumn{1}{c|}{\cellcolor[HTML]{FFFFC7}\textbf{full}} &
  \multicolumn{1}{c|}{\textbf{}} &
  \multicolumn{1}{c|}{\cellcolor[HTML]{FFFFC7}\textbf{2}} &
  \multicolumn{1}{c|}{\cellcolor[HTML]{FFFFC7}\textbf{3}} &
  \multicolumn{1}{c|}{\cellcolor[HTML]{FFFFC7}\textbf{4}} &
  \multicolumn{1}{c|}{\cellcolor[HTML]{FFFFC7}\textbf{full}} &
  \multicolumn{1}{c|}{\textbf{}} &
  \multicolumn{1}{c|}{\cellcolor[HTML]{FFFFC7}\textbf{2}} &
  \multicolumn{1}{c|}{\cellcolor[HTML]{FFFFC7}\textbf{3}} &
  \multicolumn{1}{c|}{\cellcolor[HTML]{FFFFC7}\textbf{4}} &
  \multicolumn{1}{c|}{\cellcolor[HTML]{FFFFC7}\textbf{full}} &
  \multicolumn{1}{c|}{\textbf{}} &
  \multicolumn{1}{c|}{\cellcolor[HTML]{FFFFC7}\textbf{2}} &
  \multicolumn{1}{c|}{\cellcolor[HTML]{FFFFC7}\textbf{3}} &
  \multicolumn{1}{c|}{\cellcolor[HTML]{FFFFC7}\textbf{4}} &
  \multicolumn{1}{c|}{\cellcolor[HTML]{FFFFC7}\textbf{full}} \\ \cline{1-1} \cline{3-6} \cline{8-11} \cline{13-16} \cline{18-21} 
\textbf{Dataset} &
  \textbf{} &
  \multicolumn{19}{c}{\textbf{}} \\ \cline{1-1} \cline{3-6} \cline{8-11} \cline{13-16} \cline{18-21} 
\multicolumn{1}{|l|}{\cellcolor[HTML]{FFFFC7}VC1-O (cleaned)} &
  \multicolumn{1}{l|}{} &
  \multicolumn{1}{l|}{\begin{tabular}[c]{@{}l@{}}2.94 \\ /0.2\end{tabular}} &
  \multicolumn{1}{l|}{\begin{tabular}[c]{@{}l@{}}2.94 \\ /0.2\end{tabular}} &
  \multicolumn{1}{l|}{\begin{tabular}[c]{@{}l@{}}2.94 \\ /0.2\end{tabular}} &
  \multicolumn{1}{l|}{\begin{tabular}[c]{@{}l@{}}2.94 \\ /0.2\end{tabular}} &
  \multicolumn{1}{l|}{} &
  \multicolumn{1}{l|}{\begin{tabular}[c]{@{}l@{}}1.88 \\ /0.13\end{tabular}} &
  \multicolumn{1}{l|}{\begin{tabular}[c]{@{}l@{}}1.88 \\ /0.13\end{tabular}} &
  \multicolumn{1}{l|}{\begin{tabular}[c]{@{}l@{}}1.88 \\ /0.13\end{tabular}} &
  \multicolumn{1}{l|}{\begin{tabular}[c]{@{}l@{}}1.88 \\ /0.13\end{tabular}} &
  \multicolumn{1}{l|}{} &
  \multicolumn{1}{l|}{\begin{tabular}[c]{@{}l@{}}2.37 \\ /0.16\end{tabular}} &
  \multicolumn{1}{l|}{\begin{tabular}[c]{@{}l@{}}2.37 \\ /0.16\end{tabular}} &
  \multicolumn{1}{l|}{\begin{tabular}[c]{@{}l@{}}2.37 \\ /0.16\end{tabular}} &
  \multicolumn{1}{l|}{\begin{tabular}[c]{@{}l@{}}2.37\\  /0.16\end{tabular}} &
  \multicolumn{1}{l|}{} &
  \multicolumn{1}{l|}{\textbf{\begin{tabular}[c]{@{}l@{}}1.88 \\ /0.13\end{tabular}}} &
  \multicolumn{1}{l|}{\textbf{\begin{tabular}[c]{@{}l@{}}1.88 \\ /0.13\end{tabular}}} &
  \multicolumn{1}{l|}{\textbf{\begin{tabular}[c]{@{}l@{}}1.88 \\ /0.13\end{tabular}}} &
  \multicolumn{1}{l|}{\textbf{\begin{tabular}[c]{@{}l@{}}1.88 \\ /0.13\end{tabular}}} \\ \cline{1-1} \cline{3-6} \cline{8-11} \cline{13-16} \cline{18-21} 
\multicolumn{1}{|l|}{\cellcolor[HTML]{FFFFC7}SRE'18 dev} &
  \multicolumn{1}{l|}{} &
  \multicolumn{1}{l|}{\begin{tabular}[c]{@{}l@{}}19.36 \\ /0.71\end{tabular}} &
  \multicolumn{1}{l|}{\begin{tabular}[c]{@{}l@{}}16.21 \\ /0.63\end{tabular}} &
  \multicolumn{1}{l|}{\begin{tabular}[c]{@{}l@{}}14.33\\  /0.57\end{tabular}} &
  \multicolumn{1}{l|}{\begin{tabular}[c]{@{}l@{}}8.21 \\ /0.35\end{tabular}} &
  \multicolumn{1}{l|}{} &
  \multicolumn{1}{l|}{\begin{tabular}[c]{@{}l@{}}15.7 \\ /0.63\end{tabular}} &
  \multicolumn{1}{l|}{\begin{tabular}[c]{@{}l@{}}13.78 \\ /0.56\end{tabular}} &
  \multicolumn{1}{l|}{\begin{tabular}[c]{@{}l@{}}12.64 \\ /0.53\end{tabular}} &
  \multicolumn{1}{l|}{\begin{tabular}[c]{@{}l@{}}9.29 \\ /0.39\end{tabular}} &
  \multicolumn{1}{l|}{} &
  \multicolumn{1}{l|}{\begin{tabular}[c]{@{}l@{}}17.75 \\ /0.69\end{tabular}} &
  \multicolumn{1}{l|}{\begin{tabular}[c]{@{}l@{}}15.21 \\ /0.6\end{tabular}} &
  \multicolumn{1}{l|}{\begin{tabular}[c]{@{}l@{}}13.5 \\ /0.55\end{tabular}} &
  \multicolumn{1}{l|}{\begin{tabular}[c]{@{}l@{}}8.08 \\ /0.36\end{tabular}} &
  \multicolumn{1}{l|}{} &
  \multicolumn{1}{l|}{\textbf{\begin{tabular}[c]{@{}l@{}}15.7 \\ /0.63\end{tabular}}} &
  \multicolumn{1}{l|}{\textbf{\begin{tabular}[c]{@{}l@{}}13.78 \\ /0.56\end{tabular}}} &
  \multicolumn{1}{l|}{\textbf{\begin{tabular}[c]{@{}l@{}}12.64 \\ /0.53\end{tabular}}} &
  \multicolumn{1}{l|}{\textbf{\begin{tabular}[c]{@{}l@{}}8.25 \\ /0.34\end{tabular}}} \\ \cline{1-1} \cline{3-6} \cline{8-11} \cline{13-16} \cline{18-21} 
\multicolumn{1}{|l|}{\cellcolor[HTML]{FFFFC7}SRE'16 eval} &
  \multicolumn{1}{l|}{} &
  \multicolumn{1}{l|}{\begin{tabular}[c]{@{}l@{}}20.24 \\ /0.79\end{tabular}} &
  \multicolumn{1}{l|}{\begin{tabular}[c]{@{}l@{}}17.3 \\ /0.71\end{tabular}} &
  \multicolumn{1}{l|}{\begin{tabular}[c]{@{}l@{}}15.65 \\ /0.66\end{tabular}} &
  \multicolumn{1}{l|}{\begin{tabular}[c]{@{}l@{}}10.86 \\ /0.49\end{tabular}} &
  \multicolumn{1}{l|}{} &
  \multicolumn{1}{l|}{\begin{tabular}[c]{@{}l@{}}17.59 \\ /0.82\end{tabular}} &
  \multicolumn{1}{l|}{\begin{tabular}[c]{@{}l@{}}15.82 \\ /0.78\end{tabular}} &
  \multicolumn{1}{l|}{\begin{tabular}[c]{@{}l@{}}14.87 \\ /0.75\end{tabular}} &
  \multicolumn{1}{l|}{\begin{tabular}[c]{@{}l@{}}12.31 \\ /0.67\end{tabular}} &
  \multicolumn{1}{l|}{} &
  \multicolumn{1}{l|}{\begin{tabular}[c]{@{}l@{}}18.57 \\ /0.76\end{tabular}} &
  \multicolumn{1}{l|}{\begin{tabular}[c]{@{}l@{}}16.1 \\ /0.7\end{tabular}} &
  \multicolumn{1}{l|}{\begin{tabular}[c]{@{}l@{}}14.7 \\ /0.66\end{tabular}} &
  \multicolumn{1}{l|}{\begin{tabular}[c]{@{}l@{}}10.7 \\ /0.51\end{tabular}} &
  \multicolumn{1}{l|}{} &
  \multicolumn{1}{l|}{\textbf{\begin{tabular}[c]{@{}l@{}}17.59 \\ /0.82\end{tabular}}} &
  \multicolumn{1}{l|}{\textbf{\begin{tabular}[c]{@{}l@{}}15.82 \\ /0.78\end{tabular}}} &
  \multicolumn{1}{l|}{\textbf{\begin{tabular}[c]{@{}l@{}}14.87 \\ /0.75\end{tabular}}} &
  \multicolumn{1}{l|}{\textbf{\begin{tabular}[c]{@{}l@{}}11.02 \\ /0.56\end{tabular}}} \\ \cline{1-1} \cline{3-6} \cline{8-11} \cline{13-16} \cline{18-21} 
\multicolumn{1}{|l|}{\cellcolor[HTML]{FFFFC7}SRE'19 eval} &
  \multicolumn{1}{l|}{} &
  \multicolumn{1}{l|}{\begin{tabular}[c]{@{}l@{}}17.23 \\ /0.7\end{tabular}} &
  \multicolumn{1}{l|}{\begin{tabular}[c]{@{}l@{}}14.32 \\ /0.61\end{tabular}} &
  \multicolumn{1}{l|}{\begin{tabular}[c]{@{}l@{}}12.8 \\ /0.55\end{tabular}} &
  \multicolumn{1}{l|}{\begin{tabular}[c]{@{}l@{}}7.59 \\ /0.36\end{tabular}} &
  \multicolumn{1}{l|}{} &
  \multicolumn{1}{l|}{\begin{tabular}[c]{@{}l@{}}14.07 \\ /0.6\end{tabular}} &
  \multicolumn{1}{l|}{\begin{tabular}[c]{@{}l@{}}12.17 \\ /0.54\end{tabular}} &
  \multicolumn{1}{l|}{\begin{tabular}[c]{@{}l@{}}11.18 \\ /0.5\end{tabular}} &
  \multicolumn{1}{l|}{\begin{tabular}[c]{@{}l@{}}7.89 \\ /0.39\end{tabular}} &
  \multicolumn{1}{l|}{} &
  \multicolumn{1}{l|}{\begin{tabular}[c]{@{}l@{}}15.56 \\ /0.65\end{tabular}} &
  \multicolumn{1}{l|}{\begin{tabular}[c]{@{}l@{}}13.09 \\ /0.57\end{tabular}} &
  \multicolumn{1}{l|}{\begin{tabular}[c]{@{}l@{}}11.8 \\ /0.52\end{tabular}} &
  \multicolumn{1}{l|}{\begin{tabular}[c]{@{}l@{}}7.37 \\ /0.36\end{tabular}} &
  \multicolumn{1}{l|}{} &
  \multicolumn{1}{l|}{\textbf{\begin{tabular}[c]{@{}l@{}}14.07 \\ /0.6\end{tabular}}} &
  \multicolumn{1}{l|}{\textbf{\begin{tabular}[c]{@{}l@{}}12.17 \\ /0.54\end{tabular}}} &
  \multicolumn{1}{l|}{\textbf{\begin{tabular}[c]{@{}l@{}}11.18 \\ /0.5\end{tabular}}} &
  \multicolumn{1}{l|}{\textbf{\begin{tabular}[c]{@{}l@{}}6.82 \\ /0.35\end{tabular}}} \\ \cline{1-1} \cline{3-6} \cline{8-11} \cline{13-16} \cline{18-21} 
\multicolumn{1}{|l|}{\cellcolor[HTML]{FFFFC7}rus IVR mic} &
  \multicolumn{1}{l|}{} &
  \multicolumn{1}{l|}{\begin{tabular}[c]{@{}l@{}}9.59 \\ /0.49\end{tabular}} &
  \multicolumn{1}{l|}{\begin{tabular}[c]{@{}l@{}}7.67 \\ /0.41\end{tabular}} &
  \multicolumn{1}{l|}{\begin{tabular}[c]{@{}l@{}}6.98 \\ /0.38\end{tabular}} &
  \multicolumn{1}{l|}{\begin{tabular}[c]{@{}l@{}}5.16 \\ /0.28\end{tabular}} &
  \multicolumn{1}{l|}{} &
  \multicolumn{1}{l|}{\begin{tabular}[c]{@{}l@{}}7.1 \\ /0.39\end{tabular}} &
  \multicolumn{1}{l|}{\begin{tabular}[c]{@{}l@{}}6.13 \\ /0.33\end{tabular}} &
  \multicolumn{1}{l|}{\begin{tabular}[c]{@{}l@{}}5.78 \\ /0.32\end{tabular}} &
  \multicolumn{1}{l|}{\begin{tabular}[c]{@{}l@{}}4.88 \\ /0.27\end{tabular}} &
  \multicolumn{1}{l|}{} &
  \multicolumn{1}{l|}{\begin{tabular}[c]{@{}l@{}}8.35 \\ /0.43\end{tabular}} &
  \multicolumn{1}{l|}{\begin{tabular}[c]{@{}l@{}}6.82 \\ /0.36\end{tabular}} &
  \multicolumn{1}{l|}{\begin{tabular}[c]{@{}l@{}}6.32 \\ /0.34\end{tabular}} &
  \multicolumn{1}{l|}{\begin{tabular}[c]{@{}l@{}}4.74 \\ /0.26\end{tabular}} &
  \multicolumn{1}{l|}{} &
  \multicolumn{1}{l|}{\textbf{\begin{tabular}[c]{@{}l@{}}7.1 \\ /0.39\end{tabular}}} &
  \multicolumn{1}{l|}{\textbf{\begin{tabular}[c]{@{}l@{}}6.13 \\ /0.33\end{tabular}}} &
  \multicolumn{1}{l|}{\textbf{\begin{tabular}[c]{@{}l@{}}5.78 \\ /0.32\end{tabular}}} &
  \multicolumn{1}{l|}{\textbf{\begin{tabular}[c]{@{}l@{}}4.76 \\ /0.27\end{tabular}}} \\ \cline{1-1} \cline{3-6} \cline{8-11} \cline{13-16} \cline{18-21} 
\multicolumn{1}{|l|}{\cellcolor[HTML]{FFFFC7}rus IVR phn} &
  \multicolumn{1}{l|}{} &
  \multicolumn{1}{l|}{\begin{tabular}[c]{@{}l@{}}8.04 \\ /0.46\end{tabular}} &
  \multicolumn{1}{l|}{\begin{tabular}[c]{@{}l@{}}6.53 \\ /0.38\end{tabular}} &
  \multicolumn{1}{l|}{\begin{tabular}[c]{@{}l@{}}5.86 \\ /0.34\end{tabular}} &
  \multicolumn{1}{l|}{\begin{tabular}[c]{@{}l@{}}4.05 \\ /0.24\end{tabular}} &
  \multicolumn{1}{l|}{} &
  \multicolumn{1}{l|}{\begin{tabular}[c]{@{}l@{}}6.26 \\ /0.37\end{tabular}} &
  \multicolumn{1}{l|}{\begin{tabular}[c]{@{}l@{}}5.58 \\ /0.33\end{tabular}} &
  \multicolumn{1}{l|}{\begin{tabular}[c]{@{}l@{}}5.24 \\ /0.31\end{tabular}} &
  \multicolumn{1}{l|}{\begin{tabular}[c]{@{}l@{}}4.32 \\ /0.26\end{tabular}} &
  \multicolumn{1}{l|}{} &
  \multicolumn{1}{l|}{\begin{tabular}[c]{@{}l@{}}7.11 \\ /0.41\end{tabular}} &
  \multicolumn{1}{l|}{\begin{tabular}[c]{@{}l@{}}6 \\ /0.35\end{tabular}} &
  \multicolumn{1}{l|}{\begin{tabular}[c]{@{}l@{}}5.49 \\ /0.32\end{tabular}} &
  \multicolumn{1}{l|}{\begin{tabular}[c]{@{}l@{}}4.03 \\ /0.24\end{tabular}} &
  \multicolumn{1}{l|}{} &
  \multicolumn{1}{l|}{\textbf{\begin{tabular}[c]{@{}l@{}}6.26 \\ /0.37\end{tabular}}} &
  \multicolumn{1}{l|}{\textbf{\begin{tabular}[c]{@{}l@{}}5.58 \\ /0.33\end{tabular}}} &
  \multicolumn{1}{l|}{\textbf{\begin{tabular}[c]{@{}l@{}}5.24 \\ /0.31\end{tabular}}} &
  \multicolumn{1}{l|}{\textbf{\begin{tabular}[c]{@{}l@{}}3.85 \\ /0.24\end{tabular}}} \\ \cline{1-1} \cline{3-6} \cline{8-11} \cline{13-16} \cline{18-21} 
\end{tabular}
}
}
\end{table*}

\section{Experimental setup}
\label{sec:setup}

\subsection{Train datasets}
\label{ssec:train dataset}
A wide variety of different datasets containing telephone and microphone data from proprietary datasets and from those available online were used to train the SR systems: Switchboard2 Phases 1, 2 and 3, Switchboard Cellular, Mixer 6 Speech, NIST SREs 2004 - 2010, NIST SRE 2018 (eval set), concatenated VoxCeleb (VC) 1 and 2, RusTelecom v2, RusIVR train corpus.

RusTelecom v2 is an extended version of a private Russian corpus of telephone speech, collected by call centers in Russia. 
RusIVR is a private russian corpus with telephone and media data, collected in various scenarios and recorded using different types of devices (telephone, headset, far-field microphone, etc).
In total, this training set contains 532,541 records from 33,466 speakers.

\textit{Augmentations.}
\label{sssec:augments}
During the training process online augmentation scheme was used for raw audio samples with the following settings: MUSAN additive noise with $p=0.25$, RIR convolution with $p=0.25$, frequency masking with $p=0.25$, time masking with $p=0.25$, clipping distortion with $p=0.25$.
Here $p$ is a probability of applying augmentation type for the sample in the training batch. All considered augmentations were applied in sequence.

\subsection{Considered systems}
\label{considere_systems}
During our investigation there were developed different w2v-tdnn-based systems:
\begin{itemize}
\setlength\itemsep{-0.3em}
\item \textit{w2v-tdnn-long}: the system trained on 16 seconds long speech segments on the training set (see \ref{ssec:train dataset});
\item \textit{w2v-tdnn-short}: the system trained on 2 seconds speech segments on the same train set;
\item \textit{w2v-tdnn-pooled}: the system trained on pooled 2 and 16 seconds speech segments on the same train set;
\item \textit{w2v-tdnn-uni}: the proposed universal system with two base encoders inside: w2v-tdnn-long and w2v-tdnn-short. These encoders are sharing common embedding space. During embedding extraction the encoder is selected based on input speech segment duration (see \ref{sec:uni-encoders}) using 4 seconds selection threshold.
\end{itemize}

\subsection{Evaluation data and metrics}
\label{ssec:test dataset}

The following sets were used for the systems evaluation:

\textbf{Microphone sets}: VoxCeleb1-O (VC1-O) cleaned test set \cite{nagrani2017voxceleb}, rus-IVR-mic - includes 200 speakers recorded in microphone channel and has 13K target and 1450K impostor trials.

\textbf{Telephone sets}: NIST SRE 2018 development set \cite{sre18}, NIST SRE 2016 evaluation set \cite{sadjadi20172016}, NIST SRE 2019 evaluation set \cite{sadjadi20202019}, rus-IVR-phn contains 200 speakers recorded in telephone channels and 59K target and 6300K impostor trials.

To evaluate SR systems in different speech duration conditions we prepared different testing protocols from original protocols using speech segments truncation with 2,3,4 second of pure speech.
We evaluate SR systems performance in terms of Equal Error Rates (EER) and minimum detection cost functions (minDCF) with $P_{tar}=0.05$ \cite{sadjadi20202019}.

\section{Results and discussion}
\label{sec:results}
The results of our main experiments are accumulated in Tables \ref{tab:main_table_full_enr} and \ref{tab:main_table_3sec_enr}. 
First one can see the lack of the generalization of SR encoders for long and short speech segments inspite of using huge pretrained transformers-based encoders: w2v-tdnn-short encoder significantly outperforms w2v-tdnn-long encoder for short duration protocols. For long duration protocols w2v-tdnn-long is much more better then w2v-tdnn-short. Additionally it should be noted that w2v-tdnn-pooled  does not provide optimal results not for long and for short speech segments protocols in general. In contrast w2v-tdnn-uni allows to obtain optimal results both for long and for short segments protocols in majority of the cases.

Worth mentioning that the proposed universal encoder solution can be extended to other cases of acoustic and channel mismatch conditions.

\section{Conclusions}
\label{sec:conclusion}
In this work we propose a new universal speaker embedding encoder scheme for different speech segment duration. The obtained results demonstrate the effectiveness of the proposed solution on the diverse types of the test evaluation protocols.

\bibliographystyle{IEEEbib}


\end{document}